\documentstyle[11pt,aaspp4]{article}
\begin{document}
\newcommand{\lya}{Lyman~$\alpha$}
\newcommand{\lyb}{Lyman~$\beta$}
\newcommand{\za}{$z_{\rm abs}$}
\newcommand{\ze}{$z_{\rm em}$}
\newcommand{\cmtwo}{cm$^{-2}$}
\newcommand{\nhi}{$N$(H$^0$)}
\newcommand{\nzn}{$N$(Zn$^+$)}
\newcommand{\ncr}{$N$(Cr$^+$)}
\newcommand{\degpoint}{\mbox{$^\circ\mskip-7.0mu.\,$}}
\newcommand{\halpha}{\mbox{H$\alpha$}}
\newcommand{\hbeta}{\mbox{H$\beta$}}
\newcommand{\hgamma}{\mbox{H$\gamma$}}
\newcommand{\kms}{\,km~s$^{-1}$}      % note leading thinspace
\newcommand{\minpoint}{\mbox{$'\mskip-4.7mu.\mskip0.8mu$}}
\newcommand{\mv}{\mbox{$m_{_V}$}}
\newcommand{\Mv}{\mbox{$M_{_V}$}}
\newcommand{\peryr}{\mbox{$\>\rm yr^{-1}$}}
\newcommand{\secpoint}{\mbox{$''\mskip-7.6mu.\,$}}
\newcommand{\sqdeg}{\mbox{${\rm deg}^2$}}
\newcommand{\squig}{\sim\!\!}
\newcommand{\subsun}{\mbox{$_{\twelvesy\odot}$}}
\newcommand{\et}{{\it et al.}~}

\def\ltsima{$\; \buildrel < \over \sim \;$}
\def\simlt{\lower.5ex\hbox{\ltsima}}
\def\gtsima{$\; \buildrel > \over \sim \;$}
\def\simgt{\lower.5ex\hbox{\gtsima}}
\def\arcs{$''~$}
\def\arcm{$'~$}
\def\erf{\mathop{\rm erf}}
\def\erfc{\mathop{\rm erfc}}
\title{A LARGE STRUCTURE OF GALAXIES AT REDSHIFT $Z\sim 3$ AND ITS COSMOLOGICAL IMPLICATIONS \altaffilmark{1}}
\author{\sc Charles C. Steidel\altaffilmark{2,3} and Kurt L. Adelberger}
\affil{Palomar Observatory, Caltech 105--24, Pasadena, CA 91125}
\author{\sc Mark Dickinson\altaffilmark{4,5}}
\affil{Department of Physics and Astronomy, The Johns Hopkins University, Baltimore, MD 21218}
\author{\sc Mauro Giavalisco\altaffilmark{6}}
\affil{The Carnegie Observatories, 813 Santa Barbara Street, Pasadena, CA 91101}
\author{\sc Max Pettini}
\affil{Royal Greenwich Observatory, Madingley Road, Cambridge CB3 0EZ, UK}
\author{\sc Melinda Kellogg}
\affil{Palomar Observatory, Caltech 105--24, Pasadena, CA 91125}

\altaffiltext{1}{Based in part on observations obtained at the W.M. Keck
Observatory, which is operated jointly by the California Institute of
Technology and the University of California.} 
\altaffiltext{2}{Alfred P. Sloan Foundation Fellow}
\altaffiltext{3}{NSF Young Investigator}
\altaffiltext{4}{Alan C. Davis Fellow}
\altaffiltext{5}{also Space Telescope Science Institute, 3700 San Martin Drive, Baltimore, MD 21218}
\altaffiltext{6}{Hubble Fellow}
\begin{abstract}
We report the discovery of a highly significant concentration of galaxies 
at a redshift of $\langle z \rangle = 3.090$. The structure
is evident in a redshift histogram of photometrically selected
``Lyman break'' objects in a 9\arcm\,by 18\arcm\,field
in which we have obtained 78 spectroscopic redshifts in the range $2.0 \le z \le
3.4$. The dimensions of the structure projected on the plane of the sky are at least 
11\arcm\,by 8$'$, or 14$h_{70}^{-1}$ by 10$h_{70}^{-1}$ Mpc (comoving; $\Omega_M=1)$.  
The concentration contains 15 galaxies and one faint (${\cal R} = 21.7$)
QSO.
We consider the structure in the context of a number of cosmological
models and argue that Lyman-break galaxies must be very biased
tracers of mass, with an effective bias on mass scale $M\sim 10^{15}$M$_{\sun}$
ranging from $b \sim 2$ for $\Omega_M=0.2$ to $b \simgt 6$ for $\Omega_M=1$.
In a Cold Dark Matter scenario the large bias values suggest that 
individual Lyman-break galaxies are associated
with dark halos of mass $M\sim 10^{12}$ M$_{\sun}$, reinforcing the interpretation of
these objects as the progenitors of massive galaxies at the present epoch. 
Preliminary results of spectroscopy
in additional fields suggest that such large structures are common at 
$z \sim 3$, with about one similar structure per survey field.  The implied
space density is consistent with the possibility that we are observing moderately rich
clusters of galaxies in their early non-linear evolution.
Finally,   
the spectrum of one of the QSOs discovered in our survey
($z_{em} = 3.356$) exhibits metal line absorption systems within the
3 redshift bins having
the largest number of galaxies in field, $z = 2.93$, $3.09$, and $3.28$. 
These results are the first from an ongoing ``targeted'' redshift
survey designed to explore the nature and distribution of star-forming galaxies
in the redshift range $2.7 \simlt z \simlt 3.4$.

\end{abstract}
%\newpage

\section{INTRODUCTION}

The large-scale distribution of galaxies 
at early epochs provides a powerful means
of discriminating amongst various cosmological world models
and mechanisms for the formation of structure (see, e.g., White 1996 and
references therein).  The most
comprehensive surveys of the large--scale distribution
of galaxies have been carried out in the relatively ``local'' universe 
(e.g., Shectman \et 1996), while 
hints of what is happening at larger redshifts ($z \simlt 0.8$) have
been obtained using pencil--beam apparent-magnitude limited redshift
surveys (e.g., Broadhurst \et 1990, Carlberg \et 1997, Le F\`evre
\et 1996, Connolly \et 1996, Cohen \et 1996a,b). A general result seems
to be that the (small scale) correlation function of the more distant galaxies
is reduced significantly in amplitude relative to the present
time (e.g., Efstathiou 1995, Brainerd \et 1995, Le F\'evre \et 1996, etc.), 
while on larger scales the one-dimensional, ``line--of--sight''  structures
appear to be quite prominent 
to at least $z \sim 1$, with a typical comoving distance between structures
of $\sim 100h^{-1}$ Mpc along the line of sight. It is not yet clear
how these two observational results should be combined to form
a coherent picture, since
it is so difficult to obtain information in the ``transverse''
direction in surveys of faint galaxies, so that the nature
of the structures giving rise to the ``spikes'' in one-dimensional
redshift surveys is ambiguous (Kaiser \& Peacock 1991). Locally, at least, it appears that
the line of sight structures seen in the pencil-beam surveys are likely
to be related to the cell--like ``wall--void'' geometry seen in more extensive, 
large solid-angle redshift surveys (e.g., de Lapparent \et 1986, Landy \et 1996).   

In addition to the difficulties presented by the ``geometry'' of very deep 
redshift surveys, it is not clear to what extent galaxies are
reliable tracers of mass,
particularly at high redshifts where a large 
degree of ``bias'' is expected for objects forming within massive
dark matter halos (e.g., Bardeen \et 1986, Brainerd \& Villumsen 1992, Mo \& Fukugita 1996,
Baugh \et 1997). However,  
a measurement of the bias of a class of objects at high
redshift (and the evolution of the bias with redshift)
can be used to constrain the connection between the sites of their formation
and the overall mass distribution, which can in turn be used to constrain 
models of structure and galaxy formation (e.g., Cole \& Kaiser 1989, Mo \& White 1996). 

In this paper, we present the first results
of a large survey of the galaxy distribution at $z \sim 3$. 
Efficient photometric selection of
star-forming objects at high redshift
(Steidel, Pettini, \& Hamilton 1995; Steidel \et 1996 a,b)
and subsequent spectroscopy on the W.M. Keck telescopes
allows an assessment of the growth of structure in the
galaxy distribution at significantly higher redshift than has been
previously accessible. 
In a separate paper (Giavalisco \et 1997), we
analyze the angular correlation function of the photometrically
selected $z \sim 3$ galaxy candidates in a number of fields.  
Here we focus on a single field in which we have obtained the 
most complete spectroscopic observations to date, with data analogous
to ``pencil beam'' surveys at smaller redshifts. 
Aside from probing the structures traced by
galaxies at significantly earlier epochs than has been possible
previously, working at $z\sim 3$ also has the advantage
that a reasonable angular scale for multi-object spectroscopy
on large telescopes maps onto relatively large {\it co-moving} scales;
our field size of 9\arcm\,by 18\arcm\,at $z\sim 3$
traces (transverse) structure equivalent to a field 24\arcm\,by 48\arcm\,at $z \sim 0.5$. 
As we will discuss below, the relatively large transverse co-moving scale
provides a distinct advantage in assessing the galaxy clustering properties on
scales that remain in the linear regime to the present day, allowing
for a relatively straightforward analytic treatment.  

In this paper we use the the first observations of relatively large scale structure
traced by star forming galaxies at $z \sim 3$ to estimate the amount by which this population must
be biased relative to the overall mass distribution expected for 
standard cosmological models. The measurement of this bias will allow us to
infer a corresponding dark matter halo mass scale,  and makes possible future direct
comparisons with gravitational N-body and semi-analytic simulations of early
structure formation. We will show that the implied halo mass scale of the ``Lyman break galaxies''
supports the conclusion that they likely represent the progenitors of the
massive galaxies of the present epoch (Steidel \et 1996a, Giavalisco, Steidel,
\& Machetto 1996, Mo \& Fukugita 1996, Baugh \et 1997).

\section{OBSERVATIONS}

The present work focuses on one of the fields in a
survey for $z \sim 3$ galaxies based on ground-based images in
the $Un$, $G$, and ${\cal R}$ photometric system (Steidel \& Hamilton 1993), 
which is designed to be sensitive to the Lyman break in objects having redshifts
primarily in the range $2.7 \le z \le 3.4$. A complete description of
the observations and techniques used in this survey will be
published elsewhere (Steidel \et 1997). The field under discussion
consists of two adjacent pointings of the Palomar 5m Hale telescope with
the COSMIC prime focus camera. The camera uses
a thinned, AR-coated Tektronix $2048 \times 2048$ CCD with a 
scale of 0\secpoint283 per pixel. The images
were obtained in 1995 August and 1996 August, and typically reach
1 $\sigma$ surface brightness limits of 29 (AB) mag arcsec$^{-2}$ in each
passband. The northern field
is centered on the ``SSA22'' deep redshift survey field of Cowie \et (1996), 
at $\alpha = 22^{h} 17^{m} 34.0^{s}$  $\delta = +00^{\circ}$ 15\arcm\ 01\arcs\
(J2000), and includes most of a nearby region observed as part of the
Canada-France redshift survey (Lilly \et 1995). 
The southern field is centered 525\arcs\ south of that position. 
The size of the full  
contiguous field is 17\minpoint64 in the N-S direction 
and 8\minpoint74 in the E-W direction. We refer to the two
separate pointings ``SSA22a'' and ``SSA22b'' for the northern and
southern fields, respectively.   

Spectroscopic observations of ``Lyman break'' candidates were
obtained using the Keck I telescope in 1995 October
and 1996 August, and using the Keck II telescope in its first week
of scientific observations during 1996 October. We obtained
spectra of as many candidates as possible by using 7 different
slit masks with the Low Resolution Imaging Spectrograph
(Oke \et 1995), each covering roughly 4\arcm\,by 7\arcm\,regions
and including $\sim$15--20 candidate objects per mask. Typical total exposure
times were 1.5--3 hours per mask, in separate 1800s integrations. 
The resolution of the spectra, using a 300 line mm$^{-1}$ grating, was
approximately 12 \AA, and the grating tilt was adjusted so that
the wavelength coverage for each slit included 
the 4300-7000 \AA\ range.  
Objects were assigned slits based on a number of factors, the most important
being the maximization of the number of $z\sim 3$ candidates 
on each mask. Objects were prioritized on the basis of how
unambiguously the Lyman discontinuity could be discerned from
the broad-band photometry. About two-thirds of the objects discussed
here satisfied our ``robust'' color criteria (see Steidel \et 1995), 
${\cal R} \le 25.5$, $(G-{\cal R}) \le 1.2$, and $(Un-G) \ge (G-{\cal R})+1.5$,
while the rest were selected using our ``marginal'' 
criteria, ${\cal R} \le 25.5$, $(G-{\cal R}) \le 1.2$, and
$(G-{\cal R})+1.0 \le (Un-G) \le (G-{\cal R}) +1.5$. In practice, most of the
``marginal'' candidates that have been observed spectroscopically also
have $U_n-G > 1.6$ in order to maximize the number of galaxies in the
desired redshift range, $2.7 \le z \le 3.4$. A total
of 181 objects satisfy these adopted color criteria in the survey
region, of which 113 satisfy the ``robust'' color
criteria.  Details
of the selection criteria and the resulting redshift selection function will
be presented elsewhere (Steidel \et 1997; Dickinson \et 1997).  

Approximately 80\% of the
objects assigned slits resulted in redshifts, with
78 objects having $z >2$ and 59 having $2.7 \le z \le 3.4$.   
With the exception of a small amount of contamination by
Galactic stars (approximately 5\% of the objects satisfying the color criteria
turn out to be Galactic stars, although there is essentially no contamination
by stars for objects fainter than ${\cal R}=24$), all identified color--selected
objects are at high redshift ($z >2$). 
Some of the spectra which remain unidentified may be 
objects at somewhat lower redshift than our
primary selection window, since then
the strongest spectral features would fall in regions of
much lower instrumental sensitivity; however, there is
no obvious trend with either $U_n -G$ color or apparent ${\cal R}$
magnitude for the unsuccessful redshifts.  
A higher success rate was achieved during 1996 October
due to much improved photometry and (therefore) photometric selection,
and also higher quality slit masks. 
Nine of the objects with confirmed redshifts $z >2$ (most having
$2.0 \le z \le 2.5$) were subsequently
shown not to satisfy either of the above selection categories on
the basis of the improved photometry; in addition, two $z > 2$
objects were found serendipitously on slits designed for color--selected
objects. For consistency, in the present paper we confine ourselves to 
only the 67 $z>2$ objects satisfying the adopted photometric selection 
criteria given above, for which an accurate redshift selection function has
been determined from the results of our spectroscopy in a number of different
fields.

The redshifts of the galaxies in the sample were measured from
strong absorption features in the rest--frame far--UV, mostly 
of interstellar origin (see Steidel \et 1996a, Lowenthal \et 1997) and,
when present, the Lyman $\alpha$ emission line. For spectra in which both
absorption and emission are present, the redshifts defined by Lyman $\alpha$ 
emission are systematically higher than the absorption line redshifts
by $\langle z_{{\rm Ly}\alpha} - z_{\rm abs} \rangle = 0.008 \pm 0.004 $,
or $\sim 600 \pm 300$ \kms. Based on limited near-IR spectroscopy 
of some of the high redshift galaxies (Pettini \et 1997) it appears 
that neither the emission lines nor the absorption lines actually
coincide with the true systemic redshift of the galaxy;  
the ``true'' redshift probably lies in between $z_{\rm abs}$ and $z_{{\rm Ly}\alpha}$. 
This uncertainty probably dominates over any measurement uncertainties, and
so we adopt the standard deviation of the velocity difference as the
typical uncertainty of the measured redshifts, $\sigma_v \approx 300$ \kms. 
Because most of the spectra exhibit Lyman $\alpha$ emission at
some level, whereas a smaller number of spectra exhibit only
absorption features, we have adopted the position of Lyman $\alpha$ emission
as the primary redshift indicator. 

A histogram of the $z >2$ galaxies (and the
two faint QSOs which were discovered using the same
photometric technique) in the SSA22a+b field
sample is given in Figure 1, with a bin size of $\Delta z=0.04$,
or $\Delta v \sim 3000$ \kms at $z \sim 3$. The distribution
of the objects on the plane of the sky for both the spectroscopically
confirmed and photometric candidate $z \sim 3$ samples is shown in Figure 2. 

The most striking feature in the redshift histogram is the ``spike'' of
15 objects at $z \simeq 3.1$. In fact, one of the two $z>2$ objects whose spectrum
was obtained serendipitously also falls within the same redshift bin.  The
spectra of all 16 objects in this narrow redshift range are plotted in Figure 3,
and the relative positions and photometry for the same objects summarized
in Table 1.  We will not include the serendipitous object SSA22a S1 in the 
analysis that follows, but it is intriguing that an object with strong
Lyman $\alpha$ emission (but very weak continuum) was found despite the very small
effective solid angle covered by the mask slitlets.

\section{IMPLICATIONS OF THE GALAXY OVERDENSITY}

While spectroscopy in several fields suggests that
our $U_nG{\cal R}$ selection criteria find galaxies over the
broad range $2.7 \simlt z \simlt 3.4$ (Steidel \et 1997), in this field, the one we have
sampled most densely, nearly one quarter of the redshifts lie between
$z=3.074$ and $z=3.108$.
A concentration this strong is unlikely to arise from Poisson sampling
of our selection function---using the statistical technique
described in the appendix we find that only 1 in 400 data sets generated by randomly
drawing redshifts from our selection function (Figure 1) contain
as significant a departure from Poisson expectations---and so this group of galaxies
almost certainly indicates a true peak in the density field at $z\sim 3.1$.
From Figure 1 the approximate galaxy overdensity at this redshift is
$\delta_{\rm gal} \sim 3.5$.
At first glance this seems surprisingly large:
in the local universe the rms mass fluctuation in spheres of radius $8h_{100}^{-1}$ Mpc
is $\langle(\delta M/M)^2\rangle^{1/2}\equiv\sigma_8\sim 0.6$,
and since (for $\Omega_M=1$) each bin in the histogram corresponds to a comoving volume
of $\sim 8\times 15\times 15 h_{100}^{-3}$ Mpc$^3$ 
we expect the relative mass fluctuation among bins $\sigma_{\rm mass}$
to be approximately $\sigma_8/(1+z) \simeq 0.15$.
Yet the bin at $z\sim 3.1$ contains a galaxy overdensity $\delta_{\rm gal}$
that is 25 times larger than the expected $\sigma_{\rm mass}$.  
In this section we will see what we can learn from the existence
of this single large galaxy overdensity.  The main
conclusion will be the obvious one:  Lyman-break galaxies
must be very biased tracers of mass.  A more complete analysis of the
galaxy distribution at $z\sim 3$ will be presented elsewhere.

Our approach will be straight-forward.  We will calculate a mass associated
with the spike, treat the spike as a peak in the density field smoothed
on this mass scale, and then calculate the probability of observing a peak
so high within our survey volume for three representative
cosmologies ($\Omega_M=1$, $\Omega_M=0.2$~open, and $\Omega_M=0.3$~flat).
The mass scale and peak height both depend on the 
bias parameter $b\equiv\delta_{\rm gal}/\delta_{\rm mass}$, and
we will see how large the bias must be in order to make it reasonably
probable that we would have observed the spike in our survey volume.

The first step is to calculate the mass associated with the spike.
For concreteness we will consider only the 15 objects between $z=3.074$
and $z=3.108$ to be associated with the ``spike'' structure.  This
is the interval that maximizes the significance of the spike according
to the statistical technique in the appendix,
and within it there is no evidence for substructure.
A two-sample two-dimensional
Kolmogorov-Smirnov test (Press \et (1994)) shows at the 95\% level that galaxies in 
the spike have a different areal distribution from spectroscopically
observed galaxies outside the spike, so the apparent ``edge'' of the
spike in the southern half of the field (Figure~2) is probably real.
There is not much evidence though that we have observed an edge to the
spike in any other direction, and thus we can only set
a lower limit to its transverse size of $\sim$ 9\arcm$\times$ 11\minpoint5.
For $\Omega_M=1$, this is corresponds to 
$\sim 11h_{70}^{-1}$ by $14h_{70}^{-1}$ comoving Mpc$^2$.
The effective depth of the structure for the same cosmology is
$\sim 18h_{70}^{-1}/C$ comoving Mpc, 
where $C\equiv V_{\rm apparent}/V_{\rm true}$ takes
into account the redshift space distortions caused by peculiar
velocities.  The mass associated with the structure is therefore
$M = \bar\rho V (1+\delta_m) = 4.0\times 10^{14}(1+\delta_m)/C\,\,\,h_{70}^{-1}$M$_{\sun}$,\footnote{We note that 
this mass scale is 1--2 orders of magnitude larger than the minimum mass scale
that would be derived for typical ``spikes'' in lower-redshift pencil-beam 
surveys such as those in Cohen \et 1996a,b.  The main difference is that working with a
relatively large field at $z\sim 3$ provides 
a {\it much} larger co-moving field of view and so we are able to set a correspondingly larger
lower limit to the size of structures we observe. Given our much sparser sampling density, we are probably
sensitive {\it only} to structures on relatively large angular scales;  
because the relevant mass scales are (potentially) so different,
the relationship between this
structure and those seen in other pencil-beam surveys is not
at all clear.} 
where $\bar\rho$ is the mean density of the universe and 
$\delta_m$, the mass overdensity, 
is related to the observed galaxy overdensity $\delta_{\rm gal,obs}$
through $1+b\delta_m = C(1+\delta_{\rm gal,obs})$.
Following the prescription in the appendix we estimate a galaxy overdensity
in this region of $\delta_{\rm gal, obs} = 3.6^{+1.4}_{-1.2}$.

It remains to estimate $C$.  In principle we do not even know if
$C$ is greater or less than 1:  a collapsing object will appear 
more dense in redshift space than in real space, while an object that has already 
collapsed and virialized will appear less dense in redshift space.
We will see below though that it is difficult to produce even a
moderately non-linear overdensity on this mass scale at $z\simeq 3.1$, and
so we will make the conservative assumption that we are observing
an object which is just breaking away from the Hubble expansion rather
than an object which has already collapsed.  In this case we can use the
Zeldovich approximation to correct for peculiar velocities.  In the Zeldovich
approximation each mass element maintains its linear-theory velocity, and 
as a result the density of a fluid element evolves according to
$\rho/\bar\rho\equiv 1+\delta_m = [(1-D\lambda_1)(1-D\lambda_2)(1-D\lambda_3)]^{-1}$
where the $\lambda_i$ are initial eigenvalues of the tensor of deformation and $D(t)$
is the linear-growth factor.  
High peaks in the density field are roughly isotropic (Bardeen \et 1986),
and so we can reasonably use
a simple expression for the redshift-distortion factor $C$ ({\it e.g.} Padmanabhan (1993~\S~8))
that holds when we view a collapsing object along a principal axis:
$$C = {1-D(1+f)\lambda_3 \over 1-D\lambda_3}$$
where $f\equiv d\ln D/d\ln a \simeq \Omega_M^{4/7}(z)$ (Lahav \et 1991). 
For $\lambda_1=\lambda_2=\lambda_3$ this becomes $C = 1+f-f(1+\delta_m)^{1/3}$, which
is the expression we use.  

After we have estimated the peak height $\nu$ (defined below), we
can return to this peculiar-velocity
correction and gauge how wrong our assumption of isotropy is likely to have been.
In the high-peak limit the eigenvalues of the deformation tensor have approximate
relative sizes $\lambda_1:\lambda_2:\lambda_3$
$\sim 1+1.5/\nu:1:1-1.5/\nu$ (Bond 1988).
We will see below that $\nu$ ranges from 2--5 for the parameter values we consider
(Table 2), and so the assumption of isotropy is poor in some cases.
Fortunately the peculiar-velocity correction is smallest when
the assumption is worst, and in the end our
conclusions would not be affected much by the expected
level of anisotropy.  We make a crude attempt to quantify this 
by randomly rotating a deformation tensor with eigenvalues in
the ``typical'' ratio above and calculating the
exact peculiar velocity correction for each rotation.  Even in the
worst case ($\nu\sim 2$) the $1\sigma$ spread in
$C$ is only about 10\% across the randomly rotated
ensemble.  This is small compared to the Poisson uncertainty
in $\delta_{\rm gal,obs}$ and will be neglected.

We can now calculate $M_{\rm spike}$ and $\delta_m$ for any assumed bias.
Selected values are shown in Table 2.  The final step is to calculate
whether the density field smoothed on mass scale $M_{\rm spike}$
would plausibly contain a peak of height $\delta_m$ in our survey
volume.  This will let us assess whether an assumed bias value is
consistent with the existence of the spike.  Because rms fluctuations on the
(large) mass scale $M_{\rm spike}$ are much smaller than unity,
the smoothed density field should be well approximated by linearly evolved
initial conditions, and we should be able to use linear theory to calculate
the probability of observing a peak of height $\delta_m$.  The one
complication is that nonlinear effects may have begun to accelerate
the peak's growth;  in order to analyze the spike with linear theory
we will need to estimate the linear theory height $\delta_L$ that
corresponds to the measured peak height $\delta_m$.
Spherically-symmetric collapse is one of the few cases where nonlinear
growth is quantitatively understood, and it would be convenient if we
could use spherical collapse results to correct for nonlinear growth in
our spike.  But how accurately can we model the growth of a peak 
through spherical collapse?  One issue
is whether the peak is likely to have been roughly spherical in the first place,
and we saw above this is strongly affected by its height $\nu\equiv\delta_L/\sigma$
where $\sigma$ is the rms relative mass fluctuation on the mass scale of the spike.
A second issue is whether the previous collapse of smaller perturbations within the peak
could generate non-radial motions large enough to slow down---or prevent---the
peak's collapse ({\it e.g.} Peebles 1990).  This depends on the
spectral index of density fluctuations $n$, where $|\delta_k|^2 = Ak^n$,
because (for fixed $A$) $n$ controls the level of small-scale power.  Bernardeau
(1994) argues that for $\nu\simgt 2$ and $n<-1$ typical peaks {\it will} grow at roughly the
rate predicted by the spherical collapse model, at least until $\delta_m \simeq 4$,
and this conclusion is supported by the $N$-body experiment of Thomas \& Couchman (1992).
Since in all scenarios considered the spike overdensity satisfies $\nu\simgt 2$,
and since CDM spectra on the mass scales of interest to us have effective spectral indices
$n_{\rm eff}\simlt -1$, we will adopt the spherical collapse $\delta_m\to\delta_L$
transformation as approximated by Bernardeau: $\delta_L = 1.5[1-(1+\delta_m)^{-2/3}]$.
This should at least give us a lower limit on $\delta_L$, since previous
collapse on smaller scales (which we have neglected) slows down non-linear
growth.

Table 2 shows two estimates of the number density of regions with mass overdensity $\ge\delta_L$
for each cosmology and bias value.  The first estimate
$N_{\rm Gauss}$ is the number density $n_{pk}(>\delta_L)$ of
peaks of height $\ge\delta_L$ in the Gaussian-smoothed density
field, from Bardeen \et (1986, hereafter BBKS).  For the second estimate we use
the traditional
$$N_{\rm STH} = {\bar\rho\over 2M_{\rm spike}}\biggl[1-\erf\biggl({\delta_L\over\sqrt{2}\sigma_{\rm STH}}\biggr)\biggl],$$
where $\sigma_{\rm STH}$ is the rms relative mass fluctuation in the density field
smoothed by a spherical top-hat on mass scale $M_{\rm spike}$.  Both $N_{\rm Gauss}$
and $N_{\rm STH}$ have been normalized to units of 1 over our survey volume, which we will take
to be an 18\arcm\,by 9\arcm\,rectangular field between $z=2.7$ and $z=3.4$.  If
$N\sim 1$ in this table, then we would expect to see on
average one peak with linear height $>\delta_L$ in a volume this
size, and the corresponding bias value and cosmology are roughly consistent
with the existence of the spike.  $N_{\rm Gauss}$ ($\equiv n_{pk}$) is also
presented in graphical form in Figure 4, to give an idea of the uncertainties involved.
$N_{\rm Gauss}$ and $N_{\rm STH}$ were calculated assuming a Harrison-Zeldovich ($n=1$) primordial
spectrum modified by the
BBKS adiabatic CDM transfer function with $q=k/\Gamma h$ ($\Gamma$ is
a spectral shape parameter which we will discuss more below)
and normalized so that
$$\sigma(r_0 = 8 h_{100}^{-1} {\rm Mpc}) = {\sigma_8 \over 1+z} {g(\Omega_M(z),\Omega_\Lambda(z)) \over
g(\Omega_M(0),\Omega_\Lambda(0))}$$
where $g(\Omega_M,\Omega_\Lambda)$ is the linear growth suppression factor from Carroll \et (1992)
and $\sigma_8$, the $z=0$ cluster normalization, is from Eke \et (1996).  We 
adopt cluster normalization because it is on the same length scale as our structure;
COBE measurements are on a considerably larger scale, and uncertainties in the spectral tilt $n$
make COBE measurements less constraining.
Since the normalization we have adopted is determined on roughly the
same scale as the spike, our conclusions will be insensitive to the shape of
the power spectrum.  In making Table~2 and Figure~4 we used Sugiyama's (1995) CDM
shape parameter $\Gamma \simeq \Omega_M h_{100} \exp(-\Omega_B-\Omega_B/\Omega_M)$
with $h_{100}=0.7$ and $\Omega_B = 0.024/h_{100}^2$ (Tytler \et 1996), but
our conclusions about $b$ would not change significantly for any
value of $\Gamma$ in the range $0.1<\Gamma<0.7$.

It is clear from Table~2 and Figure~4 that an overdensity similar to the one we observe
would not occur in any of these cosmologies unless Lyman-break
galaxies were very biased tracers of mass.  The minimum bias
values for $\Omega_M=1$, $\Omega_M=0.2$ open, and $\Omega_M=0.3$ flat
are approximately 6, 2, and 4 respectively.
These high values have interesting implications.
In the biasing model of
Mo \& White (1996), the mass of a dark halo $M$ can be
(implicitly) estimated from its bias $b$ and redshift $z$ through
$$\sigma(M,z) = {\delta_c \over \sqrt{(b-1)\delta_c+1}}$$
where $\sigma(M,z)$ is the rms relative mass fluctuation on scale $M$
at redshift $z$, and $\delta_c\sim 1.7$.  Because $\sigma$ is
a decreasing function of $M$, heavier halos are more highly
biased.  From Table 2, the large bias values required to explain the spike
imply Lyman-break halo masses of a few times $10^{12}$M$_{\sun}$ in
each cosmology considered.  The exact halo mass that corresponds to
a given bias is spectrum dependent, and can change by a factor of
a few for choices of $\Gamma$ different from the $h_{70}$ Sugiyama (1995)
value we have used.

In summary, then, we started out with a galaxy overdensity that naively
seemed $\sim$ 25 times larger than the expected $\sigma_{\rm mass}$.
We argued that part of this factor of 25 could be accounted for through
peculiar velocities, part through nonlinear effects, and part
through the suppression of linear growth in cosmologies with $\Omega_M<1$.
But we were still left with a galaxy overdensity several times larger
than the expected $\sigma_{\rm mass}$, and we concluded that Lyman break
galaxies are very biased tracers of mass.  A large bias is not unexpected---provided
these galaxies are massive systems.  The existence of the spike lends
further support to a halo mass scale of $\sim 10^{12}$M$_{\sun}$ for these
galaxies.

\section{CORRELATION OF PEAKS WITH QSO ABSORPTION FEATURES}

Two of the objects in the spectroscopic sample are high
redshift QSOs. One of them, called SSA22a D13, has
$z_{\rm em} = 3.083$ (${\cal R}=21.7$) and thus is part of the prominent
redshift ``spike'' at $\langle z \rangle = 3.090$, discussed at
length in \S 3. 
A second QSO, called SSA22a D14, has $z_{\rm em}=3.356$
and ${\cal R}=20.8$, and even at the coarse resolution
($\sim 12$\AA) of our survey spectra several metal line
absorption systems are evident (see Figure 5). 
There is a probable 
damped Lyman alpha system at $z_{\rm abs} = 2.937$, an optically
thick Lyman limit system at $z_{\rm abs} = 3.288$, and another
metal line system having a relatively strong Lyman $\alpha$ 
absorption line and also the C~IV $\lambda\lambda 1548, 1550$
doublet at $z_{\rm abs} = 3.094$. We note that these coincide with
the 3 strongest peaks in the redshift histogram (although, as
discussed above, only the one at $z=3.09$ is formally of
high statistical significance). We also note
that there is no confirmed Lyman break galaxy or photometrically selected candidate that 
is near enough to the sightline of SSA22a D14 to be (plausibly) responsible
for any of these absorption systems, adding to the growing number of
high redshift absorption systems that are fainter than our
typical limits for the detection of Lyman break galaxies, ${\cal R} = 25.5$ (see
Steidel \et 1995). It would clearly be of great interest to obtain a
higher resolution spectrum of SSA22a D14 so that the nature of the
absorption line systems could be better discerned.   

There are two additional metal line
systems with relatively strong C~IV doublets but weaker Lyman $\alpha$ lines at 
$z_{\rm abs} = 2.740$ and 2.801. There are no obvious features at these
redshifts in the galaxy redshift histogram, but these redshifts
are near the tail of our selection function and so the volume is not
well-sampled using our current color criteria.  

The fortuitous placement of two reasonably bright QSOs within the
SSA22 field is obviously of great interest for a comparison
of the structure seen in emission versus that seen in absorption
along the same line of sight. Although the significance
is not completely clear, the fact that all three of the
most prominent features in the redshift histogram have
counterparts in metal line absorption systems is particularly
intriguing (cf. Francis \et 1996) and suggestive of large--scale
coherence (with large covering fraction) in the overall distribution of gas and stars at high
redshift.  While suggestions of large scale structure in absorbing gas from correlations
of QSO absorption line systems (both transverse to and along the line of sight)
have been many (e.g. Jakobsen \et 1986; Sargent \& Steidel 1987; 
Steidel \& Sargent 1992; Dinshaw and Impey 1996; Williger \et 1996,
Quashnock \et 1996),  it is now possible to compile large samples of high
redshift galaxies at the same redshifts, and there is clearly a 
great deal of very fruitful work to be done in 
combining the absorption and emission techniques to obtain as complete
a picture as possible of the distribution of baryons at these early epochs. 

\section{SUMMARY}

We have presented evidence for a large structure of galaxies
at $z \simeq 3.1$ on the basis of a pronounced ``overdensity''
in a redshift histogram of photometrically selected field
galaxies, coupled with the distribution of the galaxies within
this redshift--space ``spike'' on the plane of the sky.
A relatively simple analysis of this structure (\S 3) shows that
these early star-forming galaxies must be very biased tracers of
mass, with a higher bias ($b\simgt 6$) required in standard $\Omega_M=1$
CDM than in $\Omega_M=0.2$ open CDM ($b\simgt 2$)  or in
$\Omega_M=0.3$ $\Lambda$CDM ($b \simgt 4$).
A large bias is expected for massive galaxies, and a major
conclusion of this paper is that these Lyman-break galaxies
are indeed massive systems ($M\sim 10^{12}M_{\sun}$) as we
originally claimed on different grounds (Steidel \et 1996).
A similarly large mass for these galaxies was also predicted
by Baugh \et (1997) on the basis of simple assumptions about
star and dark-halo formation, and it is encouraging that
independent estimates of these galaxies' masses should agree so well.

An interesting sidelight is that the volume of one of our
survey fields ($\sim$ few $\times 10^5$ Mpc$^3$) is well-matched to the
present-day density of X-ray selected clusters with $kT \simgt 2.5$ keV
($\sim 3 \times 10^{-6}h_{70}^3$ Mpc$^{-3}$ (Eke \et 1996)).
In an average field of this volume we would expect to see 0.3, 1.0, 1.1 structure 
that is destined to become a cluster by the present day for
$\Omega_M=1$, $\Omega_M=0.2$ open, $\Omega_M=0.3$ flat, and this
raises the possibility that structure we have found at $z\sim 3.1$ may
be an Abell cluster in its infancy.
It is on the right mass scale, and
(depending on the bias) its linear overdensity $\delta_L$ when evolved
to the present day could reach the spherical top-hat threshold
of $\delta_L \simeq 1.7$ for collapse and virialization.
Given the sampling density of the photometrically
selected candidates in the SSA22a+b field, we have probably only found $\sim 30-50$\%
of the objects in the $\langle z \rangle =3.09$ structure to the same apparent magnitude level. If
each Lyman-break galaxy is the progenitor of a present-day
galaxy brighter than $L^{\ast}$, this would imply $\sim 30-50$ such galaxies within
the structure, lending further support to the notion that the ``spike''
is a proto-cluster.  In many respects
the structure we have identified is very similar to a less evolved version of
the structure found at $z=0.985$ by Le F\'evre \et (1994) as part of the CFRS redshift survey.

The results we have presented above are based on a single field of a
``targeted'' redshift survey using Lyman break photometric
selection criteria and
focusing on the redshift range $2.7 \simlt z \simlt 3.4$.
A larger sample will test the validity of the conclusions outlined
here and will give a full picture of the correlation function of
star--forming galaxies at high redshift on scales from kpc to tens of Mpc.
Whether or not one can reach significant cosmological conclusions on
the basis of only about 70 redshifts in one field, we regard
it as extremely promising that one can now feasibly {\it observe}
the large--scale distribution of galaxies at very early epochs.
We would also like to emphasize how efficiently one can address
these issues by using photometric methods to ``pre-select'' the volume/epoch 
studied.  Lyman break galaxies at $z \sim 3$ are merely one example.

\bigskip
\bigskip

We would like to thank the staff of both the Palomar and the Keck Observatories,
as well as the entire team of people, led by J.B. Oke and J. Cohen, responsible
for the Low Resolution Imaging Spectrograph, for making these observations
possible.  We would also like to
thank A. Phillips for allowing us to use his slitmask alignment software. 
We benefited from several useful conversations with T. Padmanabhan and
J. Peacock.  S.D.M. White's detailed comments on an earlier draft improved
the clarity of \S 3. We are grateful to the referee, D. Koo, for
constructive comments. CCS acknowledges support from the U.S. National Science Foundation through
grant AST 94-57446, and from the Alfred P. Sloan Foundation. 
MG has been supported through grant HF-01071.01-94A from the Space Telescope
Science Institute, which is operated by the Association of Universities for
Research in Astronomy, Inc. under NASA contract NAS 5-26555.

\appendix

\section{STATISTICAL SIGNIFICANCE OF FEATURES IN THE REDSHIFT DISTRIBUTION}

We use a simple method to look for clustering in the galaxy redshifts.
Rather than place the redshifts into bins and look for unusually
crowded bins ({\it e.g.} Cohen \et 1996a), we scan our unbinned data
for galaxy pairs, triplets, quartets, and so on whose redshifts are closer
together than we would expect from Poisson statistics.  If the product of
our selection function and the density of Lyman-break galaxies were equal to
a known constant $\lambda$ over the redshift range of interest,
then without clustering the probability that a group
of $N$ galaxies would span a redshift interval $\Delta z$ would be given
by the the Erlangian distribution (Eadie {\et}1971)
$$p(\Delta z\,|\,N \lambda) = \lambda (\lambda\,\Delta z)^{N-2} \exp(-\lambda\,\Delta z) / (N-2)!$$
In this case it would be easy to find significantly clustered groups
of galaxies; we could simply consider each group of $N$ neighboring
galaxies in turn, and calculate the probability that in the absence of
clustering they would be
contained in a redshift interval smaller
than the observed $\Delta z_0$:
$$p(\Delta z < \Delta z_0) = \int_0^{\Delta z_0} p(\Delta z\,|\,N \lambda) d(\Delta z) = {\gamma(N-1,\Delta z_0) \over \Gamma(N-1)}$$
with $\gamma$, the (unnormalized) incomplete gamma function, given ({\it e.g.}) in
Press \et (1994).  If this probability were very close to zero, it
would suggest that the small separation of the $N$ neighbors
was inconsistent with Poisson sampling of a uniform background,
and we would consider those $N$ galaxies a cluster candidate.  (In this
section we use ``cluster'' to denote an arbitrarily small---or large---group
of galaxies whose proximity is unexpected from Poisson statistics.)

In practice the product $\lambda$ of our selection
function and the mean density
of Lyman-break galaxies is neither precisely known nor constant with
redshift, and so we proceed as follows.  First, we estimate the shape
of $\lambda(z)$ by placing the redshifts of all our
spectroscopically-confirmed marginal and robust Lyman-break galaxies
(there are 208, of which 69 are in SSA22a
or SSA22b) into bins of width $\Delta z = 0.2$, and fitting the resulting
histogram with a cubic spline (see Fig 1).  We then transform our
redshifts $z$ into a coordinate system $t$ in which $\lambda$ is roughly
constant, and the problem is
reduced to searching for significantly clustered groups of galaxies
in the midst of a constant Poisson background of unknown intensity $\lambda$.

This is just the ``on source / off source'' problem familiar from high-energy
astrophysics, where we observe $N_1$ counts during a time interval $t_1$ in
one part of the sky, and $N_2$ during $t_2$ in another, and must decide
whether there is any evidence for an intrinsically higher count rate
in region 2.  In our case we want to decide whether there is evidence
for the presence of a cluster ({\it i.e.} an elevated count rate) in the
redshift interval $t_2$ between two arbitrarily selected galaxies.  We begin
by estimating the ``background count rate'' $\lambda$ from the density of galaxies outside
this candidate cluster.  If there are $N_1$ galaxies at redshifts
less than or equal to the lowest-redshift cluster member and $N_3$ galaxies
at redshifts greater than or equal to the highest-redshift cluster member,
Bayes' Theorem gives
$$p(\lambda|t_1t_3N_1N_3) = { p(t_1t_3|\lambda N_1 N_3) p(\lambda | N_1 N_3) \over \int d\lambda' p(t_1 t_3 | \lambda' N_1N_3) p(\lambda'|N_1N_3)}$$
where $t_1$ is the interval between the lowest-redshift galaxy in our
sample and the lowest-redshift cluster member, and $t_3$ is the interval
between the highest-redshift cluster member and the highest-redshift galaxy
in our sample.  With this probabilistic distribution for $\lambda$, we can proceed
analogously to the case above where $\lambda$ was known exactly.
Since the probability that the group of $N_2$ galaxies would be contained
in {\it exactly} the observed interval $t_2$ is

$$ p(t_2\,|\,N_1N_2N_3t_1t_3) = \int d\lambda p(t_2|\lambda N_1N_2N_3t_1t_3) p(\lambda|N_1N_2N_3t_1t_3), $$
the probability that they would be contained in an interval
smaller than the observed $t_2$ is
$$p(t<t_2) = I_x(N_1+N_3-1,N_2-1) \equiv \zeta$$
where $x \equiv (t_1+t_3) / (t_1+t_2+t_3)$, $I_x$ is the incomplete
Beta function (Press \et 1994), and we have substituted the Erlangian
distribution for $p(t\,|\,N\lambda)$, assumed a uniform prior
for $\lambda$ over the interval $[0,\lambda_{max}]$ and taken the
limit $\lambda_{max} \to \infty$ (see Jaynes (1990) or Loredo (1992) for
more detailed derivations of similar results.)  This expression
for $\zeta$ lets us assess the degree to which the redshifts of
any $N_2$ adjacent galaxies are inconsistent with Poisson sampling
of a uniform background; groups of galaxies with $\zeta$ very close to 0
(or to 1) are not expected in the absence of clustering, and so by
calculating $\zeta$ for {\it all}
groups of $N_2$
adjacent galaxies in our sample, with $N_2 = 2,3,4, \dots$, we can
locate the most statistically significant clusters (or voids) in our redshift
distribution.  In practice
we restrict our attention to clusters (or voids) with $\Delta z \leq 0.2$,
since larger features are interpreted with this technique as part of
the selection function rather than of the galaxy distribution.

After using this procedure to locate candidate clusters in our
redshift distribution, we estimate the significance of each cluster
by comparing its $\zeta$ to the distribution of $\zeta$ in
simulated data sets with redshifts randomly drawn from the
smooth selection function in Fig 1.  If a candidate cluster
has a $\zeta$ so small that it appears in only 1 of 100 random
data sets, we assign that cluster a significance of 99\%.

This technique has the obvious flaws that the estimate of the
background level is not correct when there is more than one
strong spike in the data, and that we take no account of
correlations between galaxies in calculating significances.
Neither is critical in the present application, because
we have only one dominant spike, and it is on a scale
many times larger than the Lyman-break galaxy
correlation length (Giavalisco \et 1997).

Once we have identified a significant cluster we may want to associate an
overdensity $\delta$ with it though the relationship
$\delta\equiv n_{\rm cluster} / \bar n - 1$,
where $n_{\rm cluster}$ is the number density of galaxies within the cluster
and $\bar n$ is the background number density.  $\bar n$ will in general
be well determined by the many galaxies outside the cluster, but $n_{\rm cluster}$
is more problematic.  We are defining the edge of the cluster to fall {\it exactly}
on the lowest- and highest-redshift members, so should we include those two
galaxies in estimating $n_{\rm cluster}$?  In \S 3 we exclude from the
$n_{\rm cluster}$ calculation any galaxy whose position is explicitly used in
defining the cluster boundary.  There are three such galaxies---one sets the
low-redshift edge, another the high-redshift edge, and the third sets the
southern edge of the cluster (Fig 2)---and to estimate estimate $n_{\rm cluster}$
we use the formula
$$p(n\,|\,NV) = V {(nV)^{N-3}{\rm e}^{-nV} \over (N-3)!}$$
with $N=15$.

\bigskip
\newpage 
\begin{deluxetable}{lcrccrcr}
\tablewidth{0pc}
%\footnotesize
\scriptsize
\tablecaption{Objects Within the $\langle z \rangle = 3.09$ Structure}
\tablehead{
\colhead{Object} & RA\tablenotemark{a} & Dec\tablenotemark{a} & \colhead{${\cal R}$} & \colhead{$G-{\cal R}$} & \colhead{$U_n-G$} &
\colhead{Redshift\tablenotemark{b}} & \colhead{SFR\tablenotemark{c}} } 
\startdata
SSA22a C3 &     3.1 &     9.2 &   24.53 &    0.48 &   $>$2.48 &   3.074 &    12.3 \nl
SSA22b C32 &     1.0 &     6.8 &   24.89 &    0.50 &  $>$2.32 &   3.074 &    8.7 \nl
SSA22a C23 &     7.6 &    15.3 &   25.09 &    0.55 &  $>$2.17 &   3.075 &    7.3 \nl
SSA22a C14 &     2.1 &    13.5 &   25.07 &    0.79 &  $>$2.33 &   3.081 &    7.4\nl
SSA22a D13 &     3.7 &    14.5 &   21.59 &    0.35 &    2.28 &   3.083 &    QSO \nl
SSA22b D27 &     7.3 &     6.2 &   25.13 &    0.29 &    2.18 &   3.086 &    7.1\nl
SSA22a D2 &     4.8 &     9.7 &   23.39 &    0.95 &    2.61 &   3.086 &     34.6\nl
SSA22a C30 &     8.2 &    16.4 &   24.67 &    0.53 &   $>$2.36 &   3.090 &    10.7\nl
SSA22a MD31 &     6.0 &    16.3 &   23.30 &    0.44 &    1.84 &   3.090 &   38.0\nl
SSA22a MD20 &     5.9 &    12.6 &   24.15 &    0.50 &    1.67 &   3.091 &   17.3\nl
SSA22a M9 &     4.9 &    15.1 &   24.75 &    0.83 &   $>$2.06 &   3.094 &     10.0\nl
SSA22a C5 &     2.6 &     9.6 &   23.64 &    0.78 &   $>$2.93 &   3.099 &     27.5\nl
SSA22a S1\tablenotemark{d} &     1.4 &     8.9 &   26.5: &  \nodata    &  \nodata &   3.100 &
 \nodata \nl
SSA22a C19 &     7.8 &    15.0 &   24.07 &    0.99 &   $>$2.53 &   3.103 &    18.6\nl
SSA22a C8 &     6.4 &    10.7 &   24.17 &    0.47 &    $>$2.96 &   3.108 &     17.0\nl
SSA22a C9 &     4.0 &    10.9 &   24.41 &    0.50 &    $>$2.82 &   3.108 &     13.7\nl
\enddata
\tablenotetext{a}{In arc minutes, referring to the coordinate system shown in Figure 2}
\tablenotetext{b}{Defined by the position of the Lyman $\alpha$ emission line }
\tablenotetext{c}{Star formation rate estimated from the UV continuum, assuming
zero reddening, $\Omega_M=0.2$, $\Omega_\Lambda=0$, $H_0=70$ \kms Mpc$^{-1}$, and a Salpeter IMF 
evaluated from 0.1 M$_{\sun}$ to 100 M$_{\sun}$}
\tablenotetext{d}{Object found serendipitously--extremely faint in the continuum}      
\end{deluxetable}
\begin{deluxetable}{crrrrrrrrrrr}
\tablecolumns{12}
\tablewidth{0pc}
\tablecaption{Summary of Statistics for Different Cosmological Models}
\tablehead{
\colhead{}    &  \multicolumn{3}{c}{$\Omega_M=1$} &   \colhead{} &
\multicolumn{3}{c}{$\Omega_M=0.2$} & \colhead{} &
\multicolumn{3}{c}{$\Omega_M=0.3$,
$\Omega_{\Lambda}=0.7$} \\
\cline{2-4} \cline{6-8} \cline{10-12} \\
\colhead{} & \colhead{$b=4$} & \colhead{$b=6$} & \colhead{$b=8$} &
\colhead{} & \colhead{$b=1$} & \colhead{$b=1.5$} & \colhead{$b=2$} &
\colhead{} & \colhead{$b=2$} & \colhead{$b=3$} & \colhead{$b=4$}
}
\startdata
$M_{\rm halo}$\tablenotemark{a}  & 0.9   & 3    & 6    && 0.04 & 0.6   & 2  && 0.1  & 0.8 & 2    \nl
$M_{\rm spike}$\tablenotemark{b} & 8.2   & 6.9  & 6.2  && 14   & 11  & 8.9  && 14   & 11  & 9.8  \nl
$\delta_{\rm mass}$\tablenotemark{c} & 0.68  & 0.49 & 0.38 && 2.2  & 1.6  & 1.3  && 1.1  & 0.86 & 0.68 \nl
$\delta_{\rm mass,L}$\tablenotemark{d} & 0.44 & 0.35 & 0.29 && 0.80 & 0.71 & 0.64 && 0.60 & 0.51 & 0.44 \nl
$\nu_{\rm Gauss}$\tablenotemark{e}    & 4.9   & 3.7  & 3.0  && 3.2  & 2.7  & 2.3  && 4.3  & 3.4  & 2.8  \nl
$\nu_{\rm STH}$\tablenotemark{f}      & 3.6   & 2.7  & 2.1  && 2.6  & 2.2  & 1.9  && 3.3  & 2.6  & 2.2  \nl
$N_{\rm pk}$\tablenotemark{g}         & $10^{-4}$ & 0.03 & 0.20 && 0.02 & 0.09 & 0.22 && 0.001 & 0.03 & 0.14 \nl
$N_{\rm STH}$\tablenotemark{h}        & 0.003     & 0.07 & 0.33 && 0.03 & 0.14 & 0.35 && 0.006 & 0.07 & 0.23 \nl
\enddata
\tablenotetext{a}{Mass of the dark halos that harbor individual Lyman-break galaxies, derived
from the bias assuming the $n=1$, $h_{70}=1$ CDM spectrum discussed in the text.  Units
of $10^{12}$ M$_{\sun}$.}
\tablenotetext{b}{Mass scale of ``spike'' structure, in units of $10^{14}$ M$_{\sun}$.}
\tablenotetext{c}{True mass overdensity associated with spike.}
\tablenotetext{d}{Linear mass overdensity associated with spike.}
\tablenotetext{e}{$\delta_{\rm mass,L} / \sigma_{\rm Gauss}$, where $\sigma_{\rm Gauss}$ is
the rms relative mass fluctuation in the density field smoothed by a Gaussian on mass scale M.}
\tablenotetext{f}{$\delta_{\rm mass,L} / \sigma_{\rm STH}$, where $\sigma_{\rm STH}$ is
the rms relative mass fluctuation in the density field smoothed by a spherical top-hat on
mass scale M.}
\tablenotetext{g}{Expected number of peaks of height $\delta_{\rm mass,L}$ in Gaussian-smoothed
density field, per surveyed volume.  From BBKS. }
\tablenotetext{h}{Alternate estimate of expected peak number density.}
 
\end{deluxetable}

\newpage

\begin{figure}
\figurenum{1}
\plotone{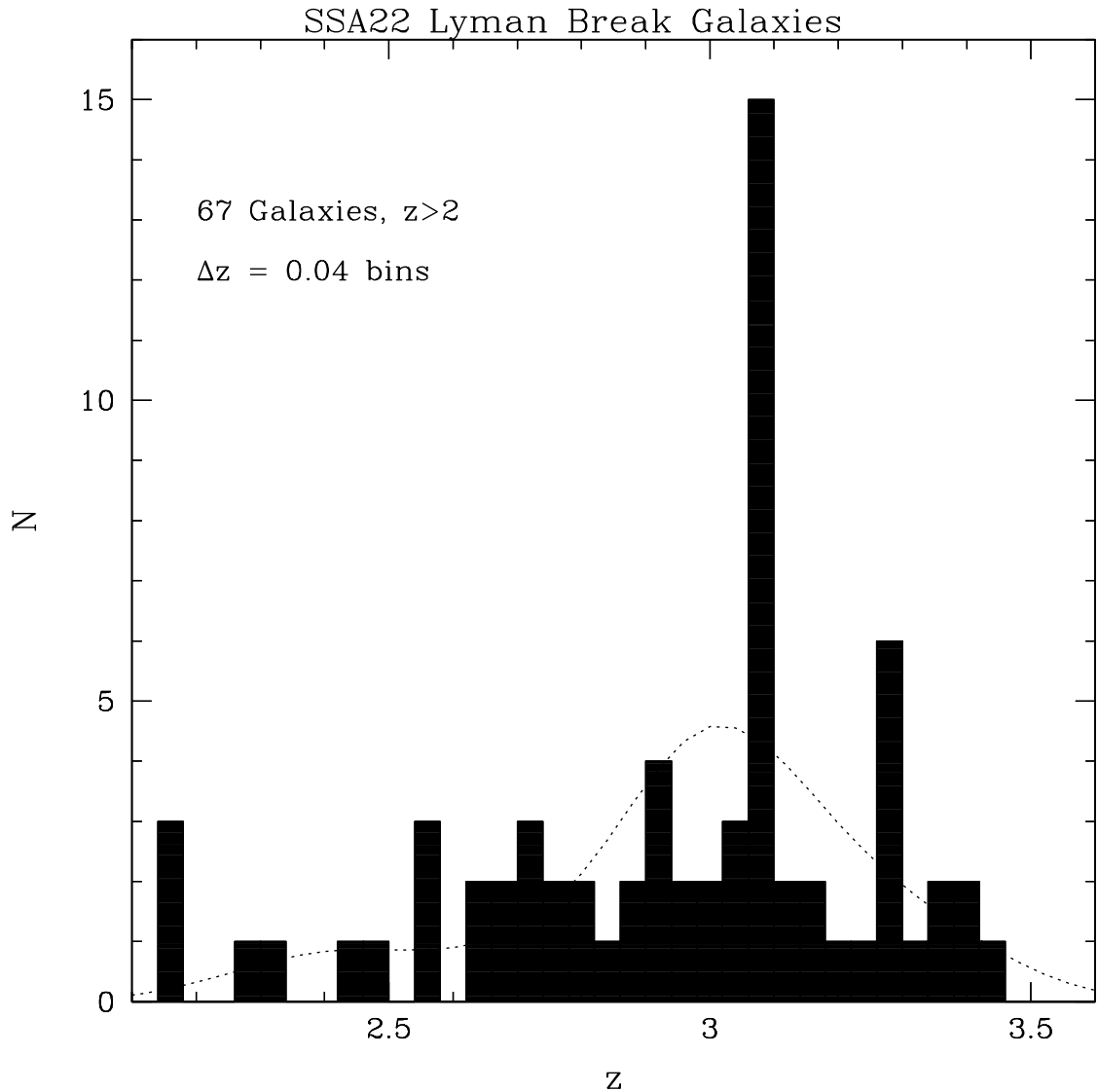}
\caption
{The redshift histogram of the 67 color--selected objects with $z >2$ that have been confirmed
spectroscopically in a 8\minpoint7 by 17\minpoint6 strip. 
The dotted curve represents the smoothed redshift selection function
obtained to date for our overall survey, normalized so as to
have the same number of total galaxies as the SSA22 sample. Approximately one-third of
the confirmed redshifts are from the SSA22 fields. The binning here is arbitrary; 
the formal boundaries of any ``features'' in the redshift distribution used for analysis
were obtained using the method described in \S 3 and in the appendix. }   
\end{figure}
\newpage
\begin{figure}
\figurenum{2}
\plotone{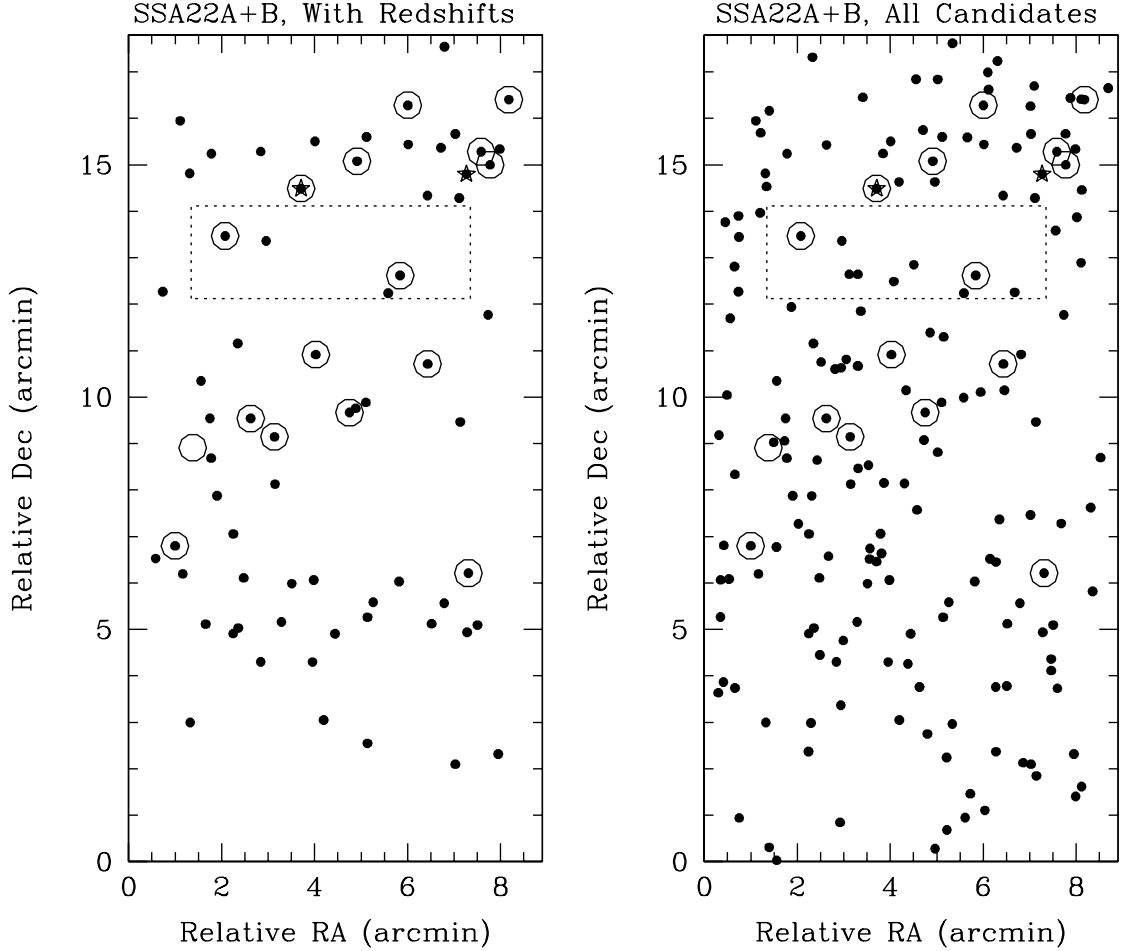}
\caption{{\it Left panel}: Distribution on the plane of the sky of all 
Lyman break objects with redshifts $z > 2$ 
The 16 objects
with $\langle z \rangle =3.090\pm 0.02$ are circled (the object with a circle but
no ``dot'' is SSA22a S1, which was found serendipitously);
the two QSOs are indicated by stars. {\it Right panel}: Same as left panel, but 
here the distribution of all color selected (using the same criteria as
for the spectroscopic sample on the left panel) candidate $z \sim 3$ galaxies on the plane
of the sky is shown. Again, objects known to be part of the structure at
$z=3.09$ are circled. 
The dotted region in both panels shows the approximate location and size
of the SSA22 Hawaii Deep Survey field (Cowie \et 1996).} 
\end{figure}
\begin{figure}
\figurenum{3}
\plotone{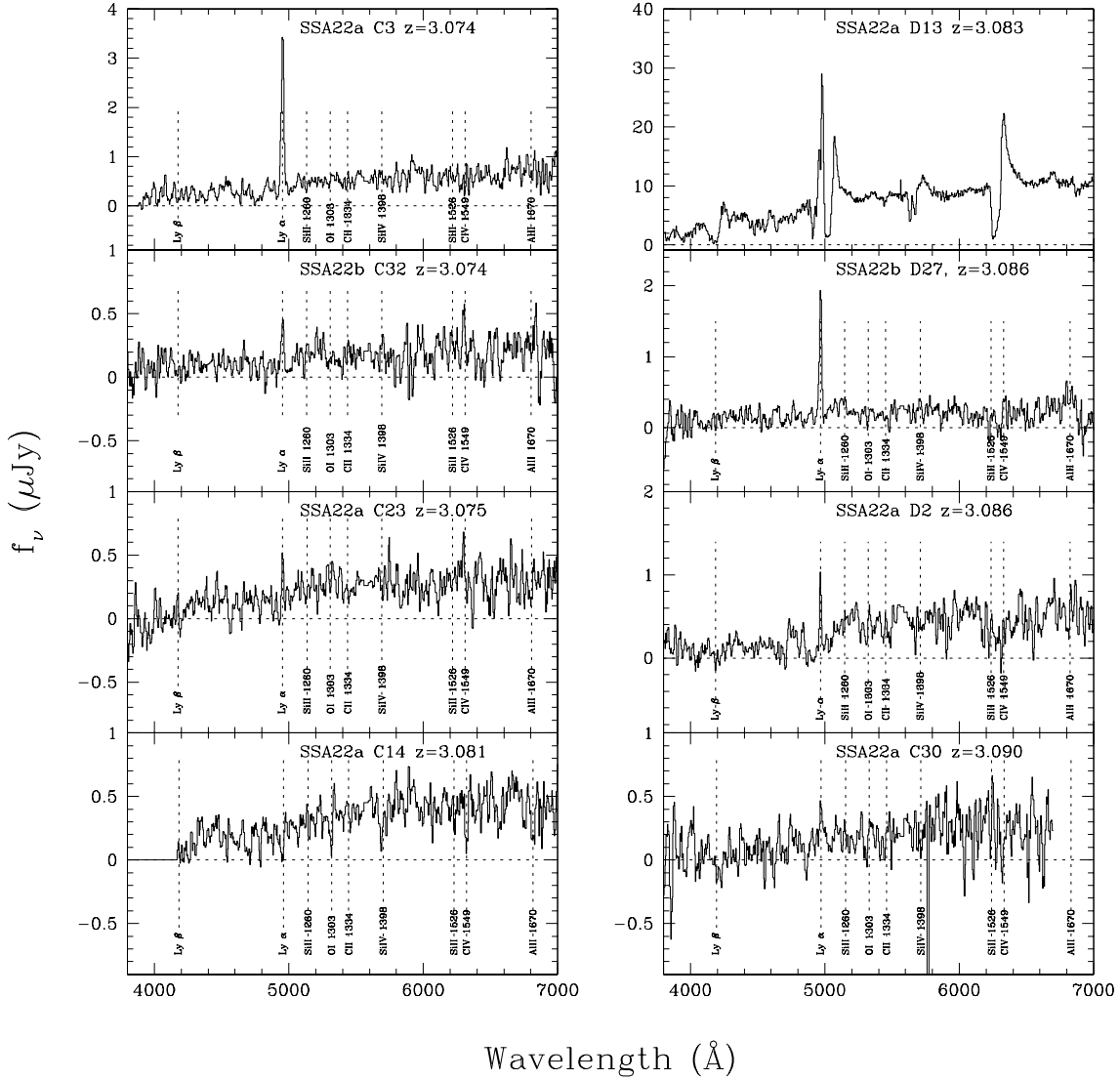}
\caption{Spectra of the 16 objects in the redshift interval $3.07 \le z \le 3.11$. 
The positions of some of the features often seen in the spectra of high-redshift 
star-forming galaxies are indicated; not all of these features are evident in
every spectrum. The absolute flux scale is only approximate; the photometry
for each object is given in Table 1. The spectra have been smoothed by a kernel
of width 12\AA\ (the spectral resolution) for display. }
\end{figure}
\begin{figure}
\figurenum{3}
\plotone{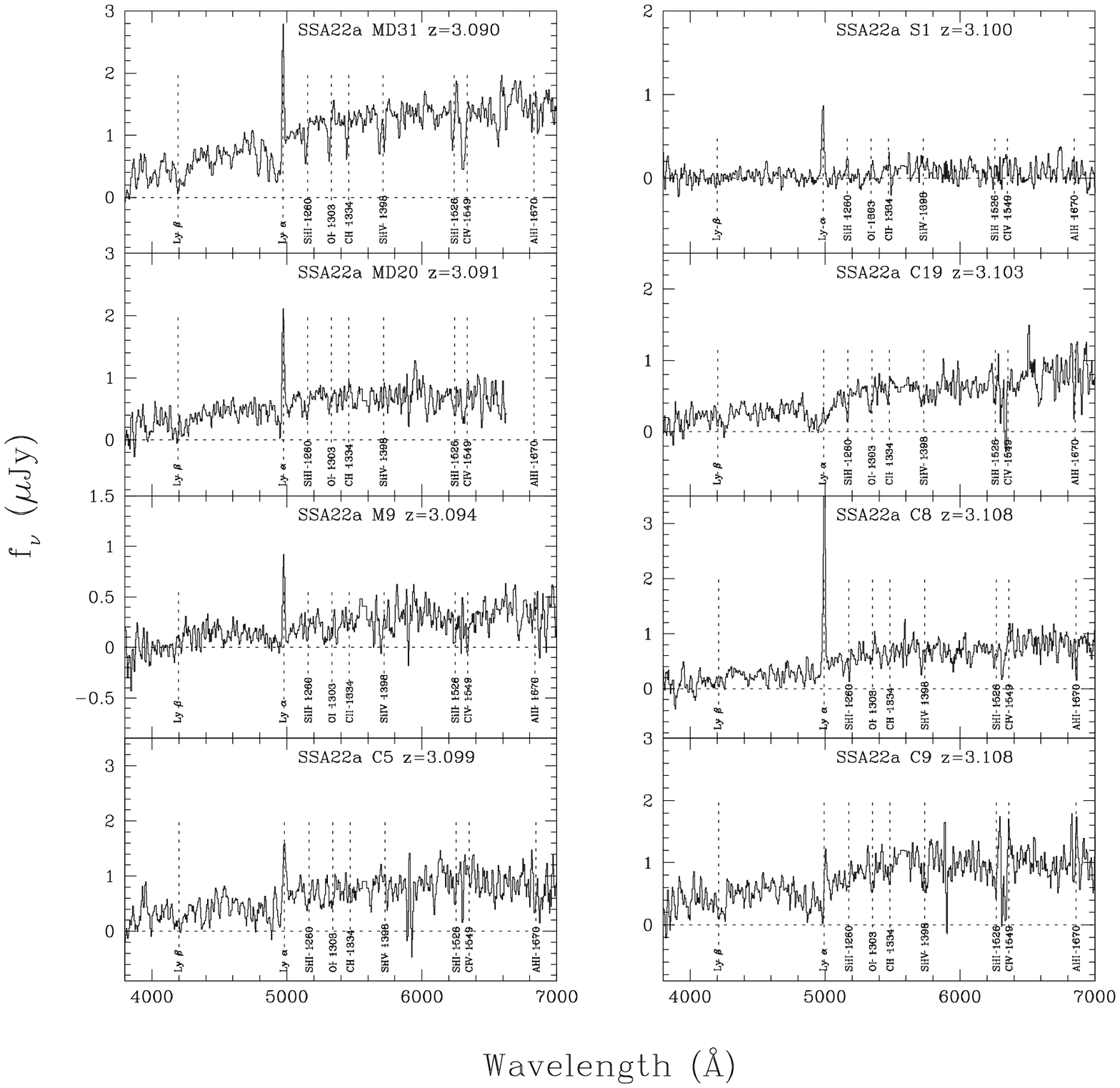}
\caption{(continued)}
\end{figure}
\begin{figure}
\figurenum{4}
%\epsscale{0.8}
\plotone{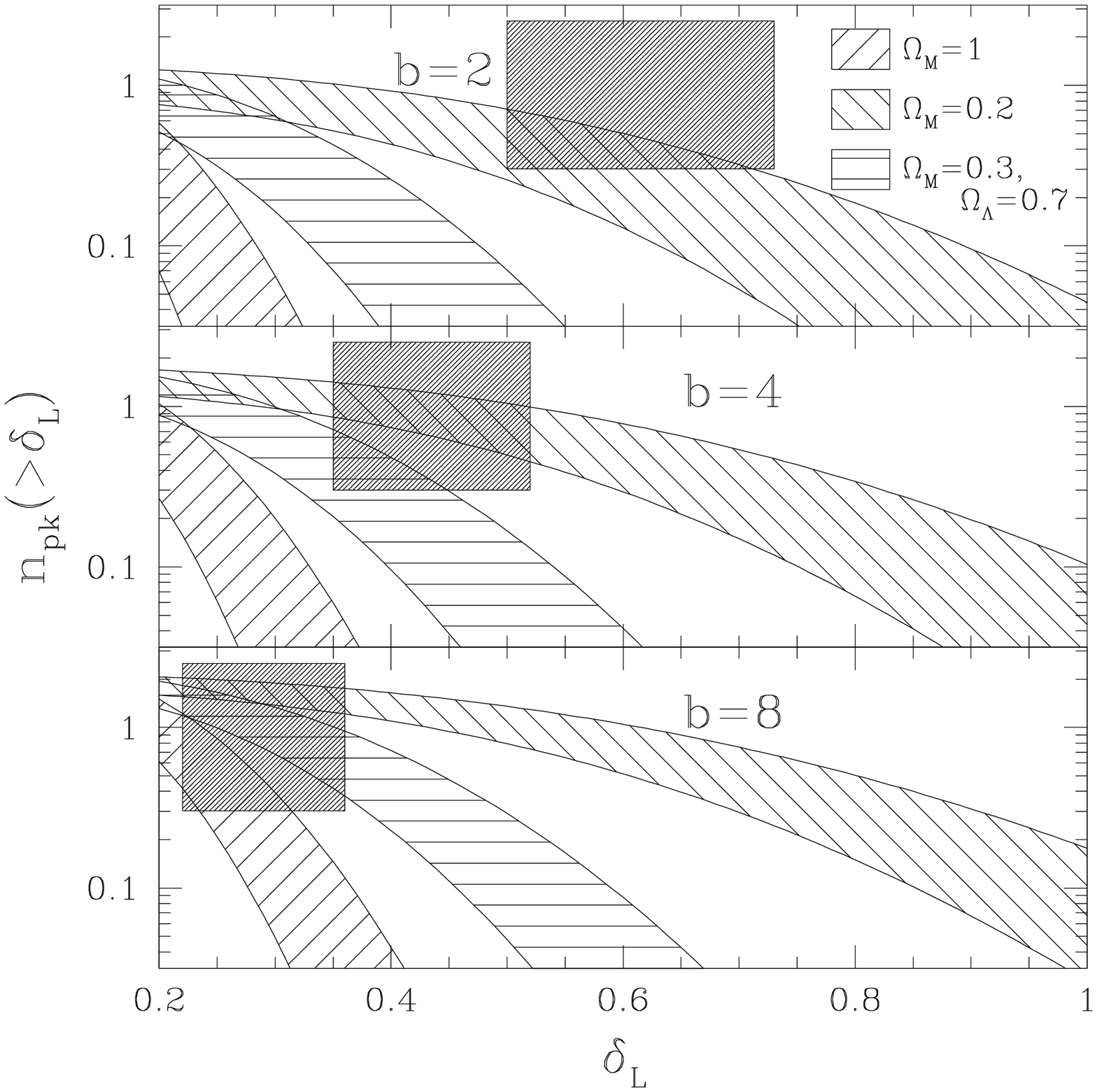}
\caption{Reconciling the observed galaxy overdensity with various cosmological scenarios.
The smooth curves show the expected number of peaks in this field with
linear overdensity greater than $\delta_L$. The shaded box shows
the approximate peak height and number density deduced from our observations.
The actual box position is slightly different for each of the cosmologies,
but this difference is insignificant for our purposes.
If a curve passes through the box then the corresponding parameter values
are at least roughly consistent with the existence of the spike in the redshift
histogram.  The $x$ range of the box is a 68\% confidence interval that takes
into account only the (Poisson) uncertainties in $\delta_{\rm gal}$; the $y$ range
of the box $0.3<n_{pk}<2.5$ is also an approximately 68\% interval on the number density
of similar structures at $z\sim3$, based on the fact that we have detected
one such overdensity in the first densely-sampled field. 
The two smooth curves for each cosmology give an idea of the uncertainty
in $n_{pk}$.  The upper curve applies if the true mass of the structure
is $1\sigma$ lower than our best estimate {\it and} the normalization
$\sigma_8$ is $1\sigma$ higher than Eke {\it et al.}'s best estimate.  The
lower curve applies if the mass of the structure is $1\sigma$ higher and
the normalization $1\sigma$ lower than the best estimates.  
See text.}
\end{figure}
\begin{figure}
\figurenum{5}
\epsscale{0.8}
\plotone{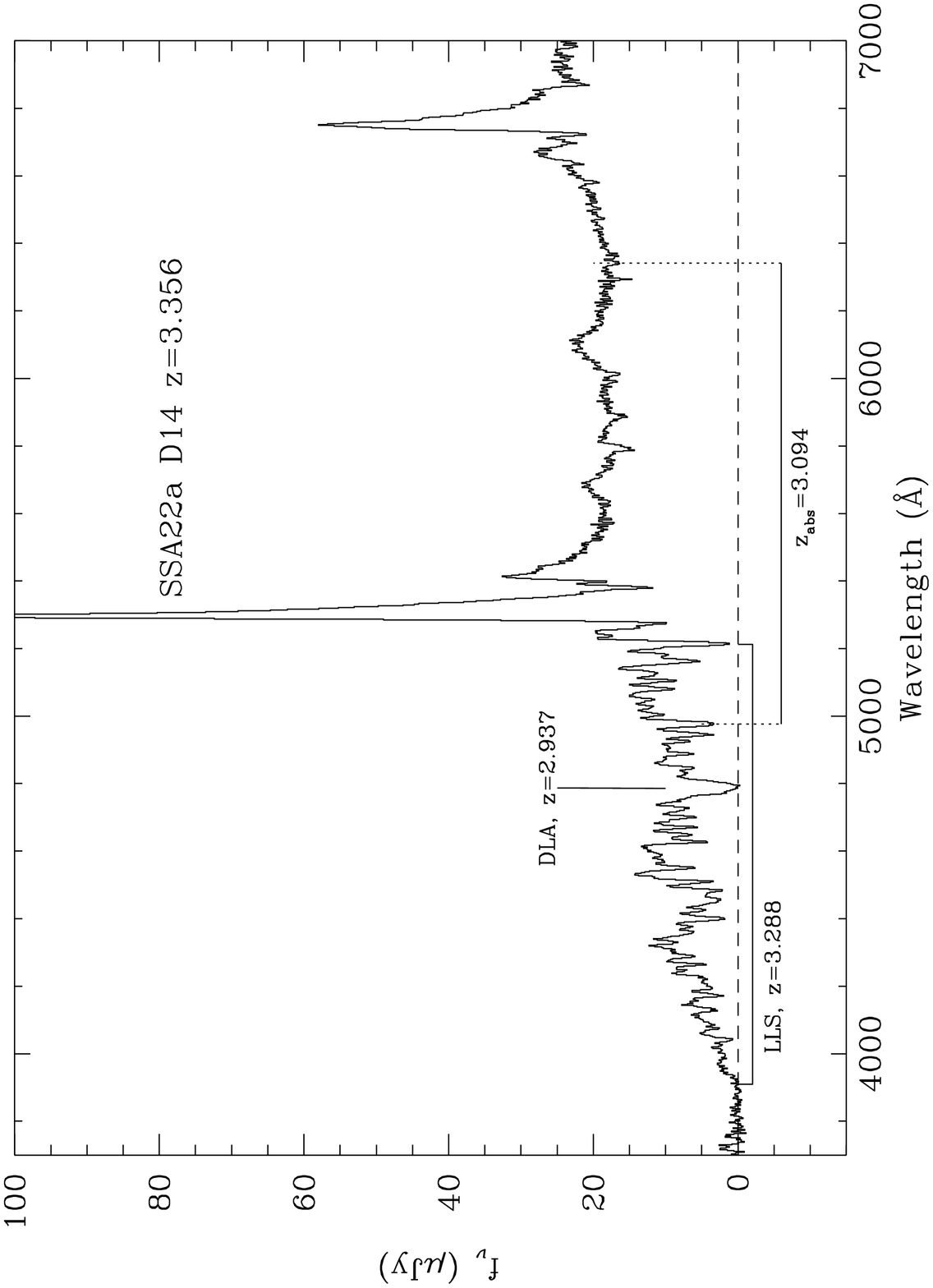}
\caption{The spectrum of QSO SSA22a D14, with the metal line absorption
systems corresponding to the redshifts of the three most
significant features in the SSA22a+b redshift histogram marked. (see text)}
\end{figure}
\end{document}